# Automated 2D and 3D Finite Element Overclosure Adjustment and Mesh Morphing Using Generalized Regression Neural Networks

Thor E. Andreassen[1], Donald R. Hume[1], Landon D. Hamilton[1], Sean E. Higinbotham[1],

Kevin B. Shelburne[1]

[1]Center for Orthopaedic Biomechanics
Mechanical and Materials Engineering
University of Denver
Denver, CO

Corresponding Author:    Thor E. Andreassen
                         Center for Orthopaedic Biomechanics
                         Mechanical and Materials Engineering
                         University of Denver
                         2155 East Wesley
                         Denver, CO 80210, USA
                         Email: thor.andreassen@du.edu
                         Phone: 303-905-7962

## Abstract

Computer representations of three-dimensional (3D) geometries are crucial for simulating systems and processes in engineering and science. In medicine, and more specifically, biomechanics and orthopaedics, obtaining and using 3D geometries is critical to many workflows. However, while many tools exist to obtain 3D geometries of organic structures, little has been done to make them usable for their intended medical purposes. Furthermore, many of the proposed tools are proprietary, limiting their use. This work introduces two novel algorithms based on Generalized Regression Neural Networks (GRNN) and 4 processes to perform mesh morphing and overclosure adjustment. These algorithms were implemented, and test cases were used to validate them against existing algorithms to demonstrate improved performance. The resulting algorithms demonstrate improvements to existing techniques based on Radial Basis Function (RBF) networks by converting to GRNN-based implementations. Implementations in MATLAB of these algorithms and the source code are publicly available at the following locations:
https://github.com/thor-andreassen/femors
https://simtk.org/projects/femors-rbf
https://www.mathworks.com/matlabcentral/fileexchange/120353-finite-element-morphing-overclosure-reduction-and-slicing

# Introduction

Computer representations of three-dimensional (3D) geometries have long been crucial for simulating systems and processes in engineering and science. In medicine, and more specifically, biomechanics and orthopaedics, obtaining and using 3D geometries is critical to many workflows, including understanding joint behavior [1]–[3], analyzing the shapes of bones and soft tissues [4], [5], finite element analysis (FEA) [6]–[10], and evaluating implant performance and surgical processes [9], [11]. Furthermore, development of robotic-assisted surgery has made 3D geometries an essential part of surgical procedures [12]. Recently, there is substantial pressure to create Digital Twins of individuals, which often relies on obtaining and using 3D structures from an individual's anatomy [13]–[15]. As such, advancements in the software tools to create, obtain, manipulate, and use 3D geometries has the potential to dramatically impact medicine and patient outcomes.

Most medical applications generate geometries of complex organic structures [16], [17] from sparse volumetric segmentation of medical images acquired with computed tomography (CT) or magnetic resonance imaging (MRI) [18], [19]. Algorithms such as Marching Cubes are used to create 3D geometries from image segmentation [20]. Semi-automated segmentation and geometry creation tools are available in many proprietary and open-source software packages [21], [22], and fully automatic tools using various algorithms [21], [22] and machine learning are a robust area of research [23], [24]. However, far less work has been done to make the resultant meshes usable in computer simulation workflows.

The limited resolution of medical images and the segmentation process result in poor-quality meshes [16], [17] comprised of nodes and elements (2D and 3D) that frequently require extensive refinement to generate a representative and useful geometry. Smoothing and remeshing are commonly applied to repair coarse, non-smooth edges; however, the methods used are rarely designed with medical applications in mind. Anatomical structures vary in deformability, size, and shape, requiring time-consuming care to repair meshes beyond the capability of existing algorithms. Moreover, most existing tools are proprietary, creating a roadblock for many. Two specific challenges in preparing meshes for computer simulation are (1) morphing to create uniformly connected meshes and (2) removing overclosures, or penetrations, between adjacent geometries.

In morphing, mesh geometries are created by non-linearly deforming or "morphing" source geometries to match a desired target geometry to efficiently create similarly connected meshes. In many applications, having different mesh geometries with identical connectivity is vital (e.g., preserving the location of ligament attachments on morphed bones). Consistent meshes are the cornerstone of shape modeling [4], [25] and are essential to the rapid creation of subject-specific finite element models [26], [27]. Morphing can also predict the location of important landmarks on the target geometry. Pillet et al. showed that ligament attachment sites of the tibia, fibula, and femur bones could be accurately predicted by morphing known locations on template bones to the geometries of new bones [28].

Many algorithms have been proposed for morphing, but the two most common are based on Coherent Point Drift (CPD) [29] or Radial Basis Functions (RBF) [30], [31]. These algorithms suffer several drawbacks. Firstly, they operate on point clouds, while most medical applications rely on 2D element surface meshes, and FEA applications rely on 3D element volumetric meshes for contact FEA [4], [32]. The performance metrics used in evaluating these algorithms are typically point-based, (e.g., Hausdorff distance [30], [33], or slice-based (e.g., Dice coefficient [30], [33], which may not capture mesh morphing performance. Secondly, these algorithms were created with the computing powers of the times in mind. Modern computing resources have the potential to improve algorithmic speed and performance. Lastly, many existing algorithms rely on prior node correspondence identification [26], [31] and, therefore, cannot be used for predicting unknown attachments and landmarks.

Penetrations, also known as overclosures, between adjacent geometries commonly occur following segmentation from CT and MRI and the smoothing and remeshing applied to fix coarse, non-smooth edges (Figure 1). Similarly, morphing high-quality template geometries onto subject-specific geometries for FEA [4], [26], [30] can create overclosures from discretization errors. In either case, overclosures between geometries in contact in FEA result in numerical instability and inaccuracy, leading to faulty predictions of deformations and unrealistic pressures between structures [32], [34].[6] Removing overclosures between organic geometries for analyses that involve contact between deformable bodies (e.g. bone, cartilage, muscle) is labor-intensive and time-consuming.

Generally, two methods to remove overclosures exist: manual node adjustment [32], [34] and automatic node adjustment [35]–[37]. Manual adjustment requires a painstaking and time-consuming process to identify and remove overclosures by manually moving nodes [32], [34]. Manual methods become infeasible as FEA models of organic geometries become more complex. Automatic algorithms provide significant time and consistency improvements over manual methods but produce jagged edges and poor mesh quality [32]. Recent developments by Idram et al. have aimed to improve automatic overclosure removal to allow geometries to fit together after 3D printing [37]. However, their method relies on flattening and smoothing surfaces, significantly changing the original topology between geometries [37]. Furthermore, many situations require unequal adjustments (i.e., preferential adjustment of one geometry over another). For example, when modeling rigid implants in contact with deformable cartilage, the implant geometry should remain unchanged while adjustments are made to the cartilage. Unequal adjustment is not possible with conventional nodal adjustment.

Recently, RBF networks have been used to solve a range of applications in the creation [38], morphing [30], [39], and repair of geometrical meshes [40]. Introduced by Broomhead and Lowe in 1988, RBF networks represent one of the earliest forms of shallow neural networks [41]. Specht created an extension of these networks for interpolation, a Generalized Regression Neural Network (GRNN), with an additional layer to allow faster training by simultaneously storing all inputs and outputs [42]. However, GRNNs have been less popular as they require far greater computing memory and power, making them impractical given past computing resources. However, with modern computing capabilities, GRNNs can increase performance over existing algorithms based on RBF networks. To address the challenge of fully automated mesh morphing and overclosure adjustment, we implemented two novel algorithms based on a Generalized Regression Neural Network (GRNN). The first algorithm (Algorithm 1) morphs a source geometry to a target geometry while allowing for dissimilar mesh densities and 2D or 3D element types. GRNN-based morphing results were compared against CPD and RBF-based morphing algorithms on a series of bony geometries. The second algorithm (Algorithm 2) removes overclosures automatically to achieve a user-desired minimum gap between a pair of meshes for 2D and 3D element types while allowing for unequal adjustment. GRNN-based overclosure removal was evaluated against a series of initially overclosed models to compare performance against conventional nodal adjustment. The proposed algorithms were implemented in MATLAB (Mathworks, Natick, MA) and made freely available online.

## Methods

The morphing and overclosure removal algorithms rely on four major processes: mesh reduction, displacement field calculation, GRNN training, and nodal calculation and adjustment. These processes are then combined into two similar but different architectures, morphing source to target geometries (Algorithm 1) and removing overclosures between pairs of geometries (Algorithm 2). The following will briefly describe each of the four processes, followed by the implementation of the four processes in the two algorithms.

### *Process 1 - Mesh Reduction*

Mesh reduction reduces the number of nodes present within each geometry, allowing both algorithms to handle high-density meshes with minimal computational overhead.
In Algorithm 2, mesh reduction creates a consistent 2D element surface for all geometries regardless of the dimensionality or mesh type (2D and 3D elements).

### *Process 2 - Displacement Field Calculation*

The displacement field calculation process calculates vectors at every node representing the optimal direction and magnitude of displacement. In Algorithm 1, the displacement field vector represents the magnitude and direction for each node to move to its nearest point in the opposite geometry. In Algorithm 2, the displacement field vector represents the magnitude and direction to achieve the desired gap between geometries.

*Process 3 - GRNN Training*

This key process involves training a GRNN to create smooth interpolations of the required adjustments. First, the entire set of nodal locations in cartesian coordinates for both geometries is concatenated into a single matrix. Second, the displacement field vectors at every node of both geometries are concatenated into a single matrix. However, prior to concatenation, the vectors of the second geometry (target geometry in Algorithm 1 and geometry 2 in Algorithm 2) are negated, allowing a single GRNN to be trained that contains all of the spatial field information in the same coordinate system for both geometries.

A GRNN is created with an input layer consisting of the three Cartesian coordinates of a point and an output layer consisting of the three predicted smooth Cartesian adjustments to apply at the node. The first hidden layer consists of a series of RBFs with different centers measuring the Euclidian 2-norm distance between points and each RBF center [43]. Within each RBF, a scalar parameter, $s$, controls the smoothness of the resulting displacement allowing for variations between large and fine adjustments. The effects of increasing or decreasing this parameter are shown in Figure 2. Increasing the smoothing allows for displacing larger regions to efficiently morph overall geometries, while decreasing the smoothing allows for localized displacements, such as for morphing finer details. Furthermore, this smoothing parameter is needed to create a

Andreassen et al. 8

GRNN that is continuously smooth, combining the displacements between geometries into one displacement field and allowing corrections of one geometry to be applied to either geometry, enabling unequal adjustment. This is shown in Figure 3, wherein the displacement field applied to the source geometry in Algorithm 1 can slowly morph to match the target geometry by creating a smoothing displacement field between the required vectors. The second hidden layer consists of summing blocks combining the RBF values and weighted values of the desired outputs. The optimum values for the weights in a GRNN are known; therefore, the model does not need to be trained using numerical optimization [42].

*Process 4 - Node Displacements*

The node displacement process uses the trained GRNN and a set of points to determine the corresponding adjustment vector approximating the required displacement field. The vectors can be used to calculate the desired displacement at any point, not necessarily those used to train the GRNN. Importantly, this allows the GRNN to be trained on a reduced version of the meshes but applied to a denser version. The new position of the points is determined by adding the resulting displacement vector to the original cartesian location of the point.

*Algorithm 1 – Morphing*

Algorithm 1 - morphing is shown in a flowchart in Figure 4 and visually after each process in Figure 5 for a 2D morphing case of a proximal tibia/fibula. The morphing algorithm uses an iterative closest point (ICP) to rigidly align the source mesh near the target mesh based on rotation and translations [44]. Then a finite-difference ICP [45], [46] adds affine transformations, allowing shearing and scaling.

The algorithm then proceeds to *Process 1*. A reduced mesh is created as a downsampled version of the original source, and target meshes with fewer nodes and elements. Several algorithms exist to perform this step for 2D triangular meshes [47] and 2D quadrilateral meshes [48]–[50]. *Process 2* uses the reduced meshes and the k-nearest neighbor (KNN) [51] or a projected distance search [27], [32], [52], [53] to determine the nearest point in the target for each point in the source using the Euclidian 2-norm distance. This is repeated for each node in the target geometry to the source geometry. The resulting displacement field defines each data point's unit vector direction and distance between the nearest target and source node. The unit vector is scaled by the distance to get the correct displacement field vector. Using the displacement field and the corresponding nodes, *Process 3* trains a GRNN following the procedure outlined above. *Process 4* then determines the desired nodal displacements to apply to the reduced source mesh and applies them. *Processes 2-4* are iterated until the reduced source mesh converges to the reduced target mesh. Following convergence, *Process 3* is repeated to train a GRNN between the original locations of the reduced source mesh and the final locations following displacement. *Process 4* is again used to determine the adjustments to apply to the original dense source mesh and applied. Lastly, *Processes 1-4* are repeated with progressively denser meshes until the dense source mesh converges to the dense target mesh within a chosen tolerance.

3D meshes can be morphed by first extracting the 2D surface of the mesh, performing all of the previous processes, and then morphing the 3D mesh as if it is another layer of dense mesh, similar to the process described in Zhang et al., in this case yielding a set of three loops, an inner one for the reduced mesh, a middle one for the dense 2D surface mesh, and an outer one for the complete 3D mesh [30]. Additionally, the final source mesh can train another GRNN network for the change in location of the dense source mesh between its initial alignment from rigid ICP to its final deformed location. This final GRNN can be helpful in many applications, including predicting locations of points near the source mesh but not directly part of it, such as attachment sites or landmarks based on the morphing of a bony geometry (Figure 6).

*Algorithm 2 – Overclosure Adjustment*

Algorithm 2 – overclosure adjustment is shown in a flowchart in Figure 7 and visually after each process in Figure 8 for an overclosure case between a gluteus maximus and gluteus medius muscle. The algorithm begins with *Process 1*, wherein two geometries of either 2D or 3D types are reduced. For 2D geometries, the density is reduced in the same manner as described for Algorithm 1 [47]–[50]. Additionally, a reduced node set comprises all nodes in the reduced mesh for each geometry. For 3D geometries, the reduced mesh is created as the connected triangular faces representing the outer surface of the 3D tetrahedral mesh or the connected quadrilateral faces representing the outer surface of the 3D hexahedral mesh. The reduced node set for 3D geometries comprises all of the nodes in the reduced mesh and a percentage of the internal nodes of the original 3D mesh chosen at random. In *Process 2*, the displacement field defines the overclosure between the reduced node sets of both target geometries. A projected distance search [27], [32], [52], [53] is performed at every node of the reduced node set onto the surface of the other reduced mesh. The Euclidian 2-norm distance between these points is calculated, with positive signs as gaps and negative signs as overclosures based on face normal directions. The desired amount of minimum gap is subtracted from each distance measure for all nodes in the reduced node set. If the distance is negative, the desired amount of gap has not been achieved,

and the magnitude of this distance scales the unit vector to get the correct direction and magnitude to remove the overclosure. If the distance is positive, the desired amount of gap has been achieved, and the displacement field vector is converted to a zero vector. The process is then repeated for the other geometry. Following the creation of the displacement field, *Process 3* trains a GRNN to approximate the displacement field at each geometry's reduced node points as previously outlined. Lastly, using *Process 4*, these displacements are applied to all of the nodes of the original dense meshes. However, since the GRNN was trained by flipping the direction of the vectors of the second geometry, any vector belonging to the second geometry must be flipped. Additionally, prior to adjustment, a single scalar parameter, $\alpha$, is applied to all the vectors, and another parameter *(1-$\alpha$)* is applied to all the other vectors for the first and second geometry, respectively, allowing for unequal adjustments between the two geometries. Following the nodal adjustment and with these new meshes as the new input meshes, the overclosures are calculated, and the algorithm stops if the desired level of the gap has been achieved; otherwise, the algorithm loops back and repeats all the previous steps with progressively denser reduced meshes.

**Validation**

Validation of each of the algorithms was done separately. For Algorithm 1, a case study of a series of different boney geometries was morphed into one another and compared against existing algorithms. For Algorithm 2, a case study of two muscles under differing levels of initial overclosure was adjusted to create a desired amount of gap and compared against conventional nodal adjustment.

*Algorithm 1 – Morphing Mesh Quality Comparison*

Sets of 4 boney geometries of the distal femur, proximal femur, full femur, and scapula were obtained from various studies and datasets [3], [54]–[56]. This resulted in 4 pairs of 2 unique 2D triangular mesh geometries from different subjects with varying mesh densities and shapes (Table 1). In each set, one geometry was randomly chosen as the source and the other as the target geometry. The GRNN algorithm was implemented to morph one geometry to the other (Figure 6 A-D). For comparison, implementations of CPD and an existing RBF-based algorithm in MATLAB were also applied [29], [57]–[59].

Special care was given to the number of iterations performed by each algorithm to compare algorithms directly. The RBF algorithm was first used to morph each geometry pair, and the clock time corresponding to a convergence tolerance of 0.01 mm was observed. Then the number of iterations of the CPD and the GRNN algorithms was changed until they reached approximately the same clock time for each geometry set.

The performance of the three algorithms was compared using five parameters: Hausdorff distance [30], surface projection distance, aspect ratio (AR) [60], dihedral face angle (DFA) [61], and total node displacement. Hausdorff distance was chosen for its prevalent use to evaluate accuracy [6], [30], [33], [37]; however, it is limited to point clouds. Surface projection distance offers insight into the accuracy of the resulting mesh surface, defined as the distance between a given point and its nearest projection point onto the other mesh [27]. AR and DFA have been linked to the performance of resulting FEA models. Lastly, the total node displacement was calculated as the distance a node moved from its starting position after the rigid ICP registration to its final position. Larger distances were assumed to mean local landmarks were moved far away from their initial locations, potentially resulting in poorly morphed geometries. Three percentile values (Median, [1$^{st}$ percentile, 99$^{th}$ percentile]) were quantified for all metrics, with lower values considered optimum.

*Algorithm 2 – Overclosure Mesh Quality Comparison*

Meshed geometries of the gluteus maximus and gluteus medius were obtained from previously published open-source models of the Visible Human Male [19], [62]–[64]. An initial overclosure was created by translating a model of the gluteus maximus into the gluteus medius . Four initial conditions were created with 0.01 mm, 0.1 mm, 1.0 mm, and 5.0 mm of overclosure (Table 2), approximating the size of overclosures observed in Andreassen et al. [19]. The GRNN algorithm and a conventional nodal adjustment algorithm were implemented in MATLAB to remove the overclosure and achieve a minimum gap of 0.01 mm. The full diagram of steps for the 5.0 mm maximum initial overclosure case is shown in Figure 8. Performance was evaluated using AR and DFA. Additionally, the change in geometry volume was calculated as a percentage change in the shape of the original meshes. An exclusive-or (XOR) operation produced a mesh containing only regions that differed between the original and adjusted geometry after overclosure removal, quantified as a percentage of the original mesh. The same percentiles reported for the morphing comparisons were used to compare the overclosure algorithms.

**Results**

*Algorithm 1 – Morphing Mesh Quality Comparison*

True clock times for each algorithm are shown in Table 3, with the RBF time as the time used as the target time for the CPD and GRNN algorithms. Additionally, all resulting quality metrics after these computation times are shown for each algorithm across each state (Table 3). All percentiles of the Hausdorff and surface projection distances across all four geometry pairs for the CPD algorithm were greater than the corresponding value for the GRNN and the RBF algorithms (Table 3). Additionally, all percentiles of the AR were greater for the RBF compared with the GRNN algorithm (Figure 9 and Table 3). The median, $1^{st,}$ and $99^{th}$ percentiles of the DFA were greater in half of the geometry pairs for the RBF algorithm compared with the GRNN and smaller for the other half. Lastly, the node movement for the RBF algorithm was always greater than the GRNN network and greater than the CPD algorithm in all but the scapula geometry case (Table 3). Critically, the CPD algorithm failed to finish a single iteration for the full femur case in the desired time of about 640 seconds (a single iteration of the algorithm took more than 4700 seconds).

*Algorithm 2 – Overclosure Mesh Quality Comparison*

AR and DFA changed by less than 0.1% across all trials (Table 4). The only exception occurred during the 5.0 mm overclosure for the conventional nodal adjustment, which failed to converge and had values of 4.87 and 46.54 for the AR and DFA, respectively, for the $99^{th}$ percentile of values. The volumetric changes increased with increased initial overclosure for both algorithms. The GRNN algorithm also had greater volumetric changes than the nodal adjustment method for all cases except the 5.0 mm case, where the nodal adjustment had a volumetric change of 22.6%.

## Discussion

This work introduced two novel algorithms to automatically morph sets of geometries (Algorithm 1) and remove overclosures (Algorithm 2) without user involvement. These algorithms were based on four major processes, including a GRNN, and implemented in MATLAB. Several case studies were created that illustrated the improved performance of the proposed algorithms against existing methods.

The GRNN algorithm may have improved performance over existing methods when the quality of the resulting meshes is crucial. When compared, the RBF algorithm and the GRNN algorithm (Algorithm 1) successfully morphed all meshes, despite differences in mesh density between the targets and source. However, as shown by the proximal tibia and fibula morphing case, the RBF method created a mesh with poor quality and nodes significantly further from their starting positions than the GRNN algorithm (Figure 9), despite both matching the point cloud of the intended target. In contrast, the CPD algorithm failed to converge in the given time with the resultant mesh very similar to the original source mesh. Furthermore, unlike RBF and CPD, Algorithm 1 can predict landmarks (Figure 6 A-D), attachment sites (Figure 6 A-D), and entire geometries (Figure 6 E), allowing for the rapid creation of subject-specific models.

As the severity of initial overclosures increased, the GRNN algorithm produced more predictable and robust results than conventional nodal adjustment. For an initially overclosed gluteus maximus and gluteus medius muscle, the GRNN algorithm displaced a larger region of nodes to create the adjustment than the conventional nodal adjustment and displaced larger regions as the severity of initial overclosure increased (Table 4). Both algorithms resulted in similar AR and DFA, demonstrating that their behavior was similar for 2D triangular meshes with small initial overclosures. However, in the case of the 5.0 mm overclosure, the conventional nodal adjustment algorithm failed to converge and resulted in poor-quality meshes that would be unsuitable for most use cases. Larger overclosures, greater than 3.0 mm, can occur in datasets with many interfacing geometries [19] making the nodal adjustment method unsuitable for these cases.

A limitation of the analysis was that the GRNN algorithms were not compared with many alternative means of mesh adjustment. Publicly available morphing implementations are limited, and of those, most are wrapped into toolboxes, with code rarely provided. As such, fair comparisons between algorithms are difficult. The implementations used here for comparison were based on common methodologies and implemented similarly in MATLAB, allowing for fair comparison. Still, the authors acknowledge that other algorithms and implementations may exist with superior performance. Similarly, the conventional nodal adjustment implementation used herein can be improved by removing the overclosure, smoothing the meshes, and then iterating. This process can be more straightforward than Algorithm 2, converge faster, and may be more readily available. However, such smoothing algorithms are often limited to 2D elements, as few algorithms can smooth 3D elements, particularly for hexahedral meshes. Furthermore, smoothing algorithms often result in undesirable volumetric changes to the meshes in regions where no overclosure adjustment is necessary. In contrast, the proposed GRNN algorithm directs the smoothing to the adjustments applied to the mesh. As a result, volumetric changes are kept to a minimum, and regions distant from the initial overclosures are unchanged. In addition, nodes within the 3D meshes that are not initially overclosed are adjusted through the smooth adjustment vectors created by the GRNN, improving AR compared with conventional nodal adjustment.

Another limitation is that the cases used for comparison represent only a small number of the possible scenarios and metrics that could be tested. The cases used were meant to represent realistic geometries and scenarios that would readily appear in existing biomechanics and orthopedics applications. Still, the authors acknowledge that many additional situations could exist where the proposed algorithms may suffer compared to other available tools. Even so, as part of other work, the authors have used the overclosure algorithm (Algorithm 2) to create a publicly available dataset of 260 unique geometries from the Visible Human Male and Female [19]. As described in Andreassen et al., 471 pairs of overclosed geometries were adjusted to create FEA-ready meshes for all segmented geometries. The overclosures ranged in size between 0.001 mm and approximately 10.0 mm [19]. The algorithm removed these overclosures while leaving all bones rigid and deforming all other geometries, demonstrating algorithm robustness (Figure 10). In addition, the morphing algorithm (Algorithm 1) is part of several ongoing studies, including an implementation to morph 3D tetrahedral meshes for use in statistical shape modeling of knees. All pseudocode for algorithms, as well as geometries, results for algorithm comparisons, validation, tests, and FEA analysis comparisons for morphing (Algorithm 1) and overclosure removal (Algorithm 2) are available as Supplemental Material; cases involving 3D elements are highlighted.

This work introduced two novel algorithms using GRNNs to morph geometries and remove initial overclosures between meshes. These algorithms are particularly useful in models resulting from sparse volumetric data. The performance of these algorithms was compared against existing methods and demonstrated improved performance. The proposed algorithms could be an alternative to existing algorithms in various medical applications, including computational frameworks for subject-specific modeling. Lastly, this work highlights that GRNNs could replace existing RBF-based algorithms to leverage more robust computing resources and simultaneously that simple algorithmic steps can be swapped with one another to create different processes for manipulating geometries. The algorithms have been implemented in MATLAB, and distributions of these algorithms have been made publicly available at:

    https://github.com/thor-andreassen/femors

    https://simtk.org/projects/femors-rbf

    https://www.mathworks.com/matlabcentral/fileexchange/120353-finite-element-morphing-overclosure-reduction-and-slicing

## Funding

NIH National Institute of Arthritis and Musculoskeletal and Skin Diseases, National Institute of Biomedical Imaging and Bioengineering, and the National Institute of Child Health and Human Development (Grant U01 AR072989).

## Competing interests

None declared

## Ethical approval

Not required

# TABLES

Table 1: Parameters for initial geometries for mesh morphing cases

|  |  | Number of Nodes | Number of Elements | Aspect Ratio | Dihedral Face Angle (deg) |
|---|---|---|---|---|---|
| Distal Femur | Source | 14524 | 29044 | 1.20 [1.02, 1.93] | 1.69 [0.00, 12.25] |
|  | Target | 6144 | 12284 | 1.29 [1.03, 2.42] | 3.98 [0.06, 43.10] |
| Tibia/Fibula | Source | 22933 | 45862 | 1.29 [1.03, 2.79] | 1.91 [0.00, 18.98] |
|  | Target | 4049 | 8094 | 1.12 [1.01, 1.52] | 3.52 [0.04, 37.05] |
| Full Femur | Source | 71700 | 143396 | 1.25 [1.03, 1.83] | 2.57 [0.03, 17.59] |
|  | Target | 71073 | 142142 | 1.49 [1.05, 2.87] | 2.59 [0.03, 17.21] |
| Scapula | Source | 35089 | 70232 | 1.11 [1.00, 2.12] | 0.83 [0.00, 74.66] |
|  | Target | 31524 | 62359 | 1.18 [1.02, 1.72] | 1.17 [0.00, 32.32] |

Table 2: Parameters for initial geometries for Overclosure Removal for different cases

|  |  | **Gluteus Maximus** 65638 Elements 32827 Nodes | **Gluteus Medius** 14592 Elements 7298 Nodes |
|---|---|---|---|
| 0.01 mm Overclosure | Number of Overclosures | 2 | 1 |
|  | % of Nodes Overclosed | 0.01% | 0.01% |
| 0.1 mm Overclosure | Number of Overclosures | 5 | 3 |
|  | % of Nodes Overclosed | 0.02% | 0.04% |
| 1.0 mm Overclosure | Number of Overclosures | 1196 | 513 |
|  | % of Nodes Overclosed | 3.64% | 7.03% |
| 5.0 mm Overclosure | Number of Overclosures | 3342 | 1404 |
|  | % of Nodes Overclosed | 10.18% | 19.24% |

Table 3: Mesh quality metrics of CPD, RBF, and GRNN morphing algorithms on different morphing cases

|  |  | CPD | RBF | GRNN |
|---|---|---|---|---|
| Distal Femur | Computation Time (s) | 398 | 387 | 422 |
|  | Hausdorff (mm) | 2.21 [0.20, 9.02] | 0.29 [0.05, 0.49] | 0.28 [0.06, 0.54] |
|  | Surface Projection (mm) | 2.20 [0.04, 9.02] | 0.00 [0.00, 0.14] | 0.05 [0.00, 0.40] |
|  | Aspect Ratio | 1.23 [1.03, 1.97] | 1.75 [1.07, 8.08] | 1.42 [1.04, 3.60] |
|  | Dihedral Face Angle (deg) | 1.73 [0.02, 12.28] | 1.97 [0.03, 149.23] | 2.19 [0.04, 14.88] |
|  | Node Movement (mm) | 0.76 [0.49, 1.14] | 24.52 [3.04, 41.94] | 4.04 [0.99, 15.45] |
| Tibia/Fibula | Computation Time (s) | 554 | 400 | 368 |
|  | Hausdorff (mm) | 2.94 [0.52, 10.73] | 0.97 [0.26, 1.43] | 0.90 [0.17, 1.50] |
|  | Surface Projection (mm) | 2.79 [0.06, 10.70] | 0.01 [0.00, 0.20] | 0.08 [0.00, 0.63] |
|  | Aspect Ratio | 1.29 [1.03, 2.79] | 2.24 [1.12, 26.59] | 1.49 [1.05, 3.83] |
|  | Dihedral Face Angle (deg) | 1.91 [0.00, 19.02] | 4.93 [0.06, 178.46] | 2.29 [0.03, 24.31] |
|  | Node Movement (mm) | 2.09 [2.09, 2.09] | 7.14 [1.58, 18.56] | 4.92 [1.04, 12.75] |
| Full Femur | Computation Time (s) | 4765* | 643 | 626 |
|  | Hausdorff (mm) | 1.93 [0.05, 6.59]* | 0.44 [0.07, 0.81] | 0.64 [0.16, 2.07] |
|  | Surface Projection (mm) | 1.99 [0.27, 6.61]* | 0.02 [0.00, 0.18] | 0.42 [0.01, 2.02] |
|  | Aspect Ratio | 1.28 [1.03, 1.90]* | 1.37 [1.04, 2.41] | 1.33 [1.03, 2.14] |
|  | Dihedral Face Angle (deg) | 2.47 [0.02, 17.73]* | 2.28 [0.04, 15.99] | 2.61 [0.03, 18.08] |
|  | Node Movement (mm) | 1.85 [1.84, 1.85]* | 5.77 [1.77, 10.60] | 3.40 [0.79, 7.29] |
| Scapula | Computation Time (s) | 967 | 577 | 537 |
|  | Hausdorff (mm) | 4.55 [0.41, 22.19] | 0.71 [0.10, 1.14] | 0.75 [0.13, 2.47] |
|  | Surface Projection (mm) | 4.50 [0.09, 22.19] | 0.02 [0.000, 0.37] | 0.21 [0.00, 2.37] |
|  | Aspect Ratio | 1.11 [1.01, 2.13] | 2.12 [1.12, 12.04] | 1.33 [1.03, 3.32] |
|  | Dihedral Face Angle (deg) | 0.84 [0.01, 77.49] | 4.59 [0.06, 174.62] | 1.73 [0.03, 69.17] |
|  | Node Movement (mm) | 29.08 [28.37, 29.53] | 10.35 [2.18, 23.38] | 5.58 [0.82, 13.28] |

Table 4: Mesh quality metrics of traditional nodal adjustment and the GRNN overclosure removal algorithms on different overclosure cases

|  |  | **NODAL** | **GRNN** |
|---|---|---|---|
| Original | Aspect Ratio | 1.39 [1.03, 2.37] | 1.39 [1.03, 2.37] |
|  | Dihedral Face Angle (deg) | 1.12 [0.02, 16.89] | 1.12 [0.02, 16.89] |
|  | Volume Change | N/A | N/A |
| 0.01 mm Overclosure | Completed/Failed | Completed | Completed |
|  | Aspect Ratio | 1.39 [1.03, 2.37] | 1.39 [1.03, 2.37] |
|  | Dihedral Face Angle (deg) | 1.13 [0.02, 16.89] | 1.12 [0.02, 16.89] |
|  | Volume Change | 0.00% | 0.00% |
| 0.1 mm Overclosure | Completed/Failed | Completed | Completed |
|  | Aspect Ratio | 1.40 [1.03, 2.38] | 1.40 [1.0324, 2.38] |
|  | Dihedral Face Angle (deg) | 1.13 [0.02, 16.89] | 1.12 [0.02, 16.89] |
|  | Volume Change | 0.00% | 0.88% |
| 1.0 mm Overclosure | Completed/Failed | Completed | Completed |
|  | Aspect Ratio | 1.40 [1.03, 2.38] | 1.40 [1.03, 2.38] |
|  | Dihedral Face Angle (deg) | 1.15 [0.02, 17.24] | 1.12 [0.02, 16.90] |
|  | Volume Change | 0.06% | 0.97% |
| 5.0 mm Overclosure | Completed/Failed | Failed | Completed |
|  | Aspect Ratio | NaN [1.03, 4.87] | 1.40 [1.03, 2.39] |
|  | Dihedral Face Angle (deg) | NaN [0.02, 46.55] | 1.09 [0.02, 17.07] |
|  | Volume Change | 22.69% | 5.91% |

# Figures

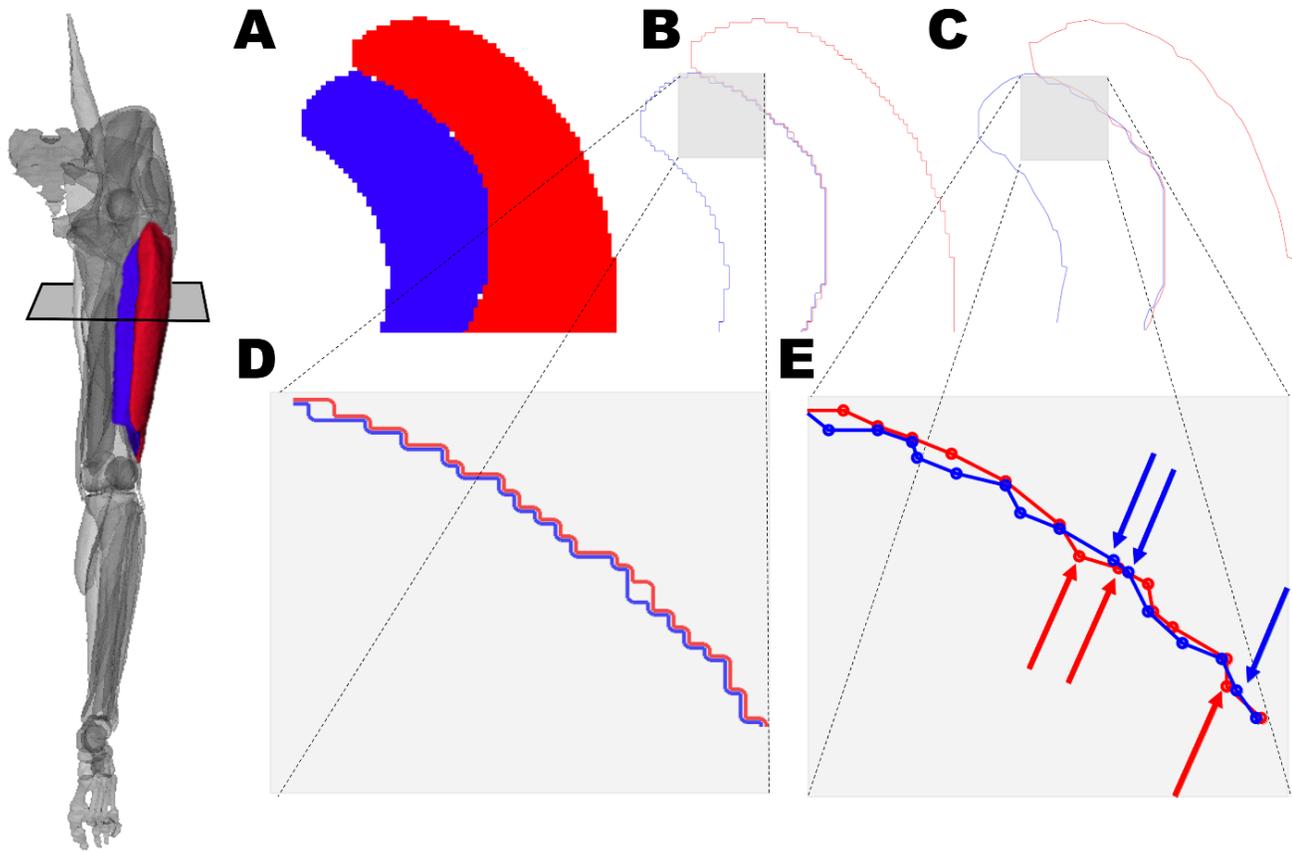

Figure 1: A diagram showing a series of steps from the original segmentation geometry to the resulting overclosures in mesh geometry. (A) Axial slice through original segmentation of vastus intermedius and vastus lateralis (B) Resulting mesh edges from Marching Cubes algorithm. (C) Resampled mesh to remove blocky edges and reduce node count. (D) Magnified region of the gray box region in B. (E) Magnified region of the gray box region in C highlighting the presence of overclosures and arrows pointing to the overclosed nodes and directions to move the node to remove overclosure.

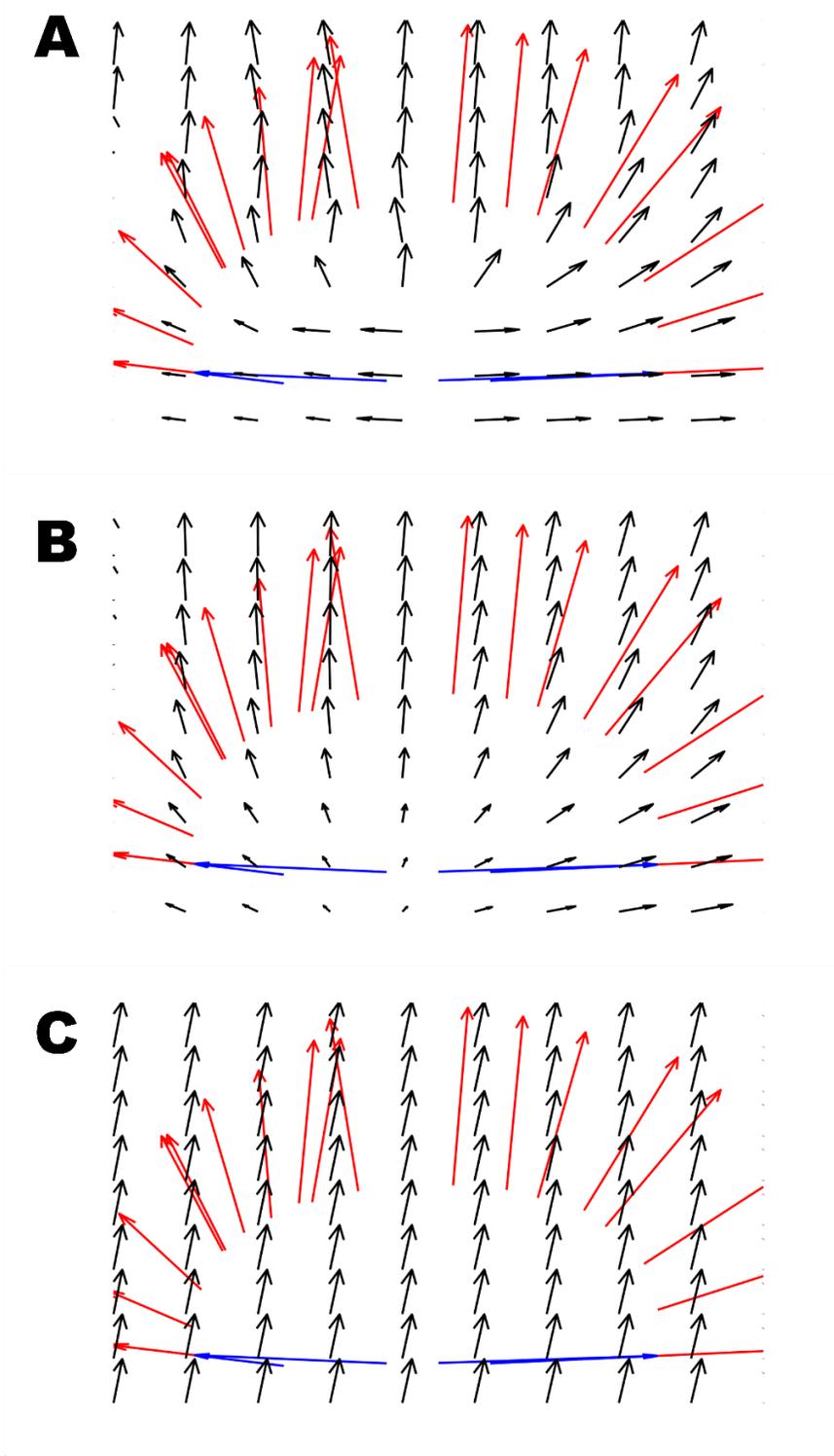

Figure 2: Simplified 2D example images showing input source vectors (blue) and flipped target vectors (red) and the resulting deformation field (black arrows) from GRNN training with three levels of smoothing (A) low smoothing $s=1$ (B) medium smoothing $s=10$ (C) high smoothing $s=100$

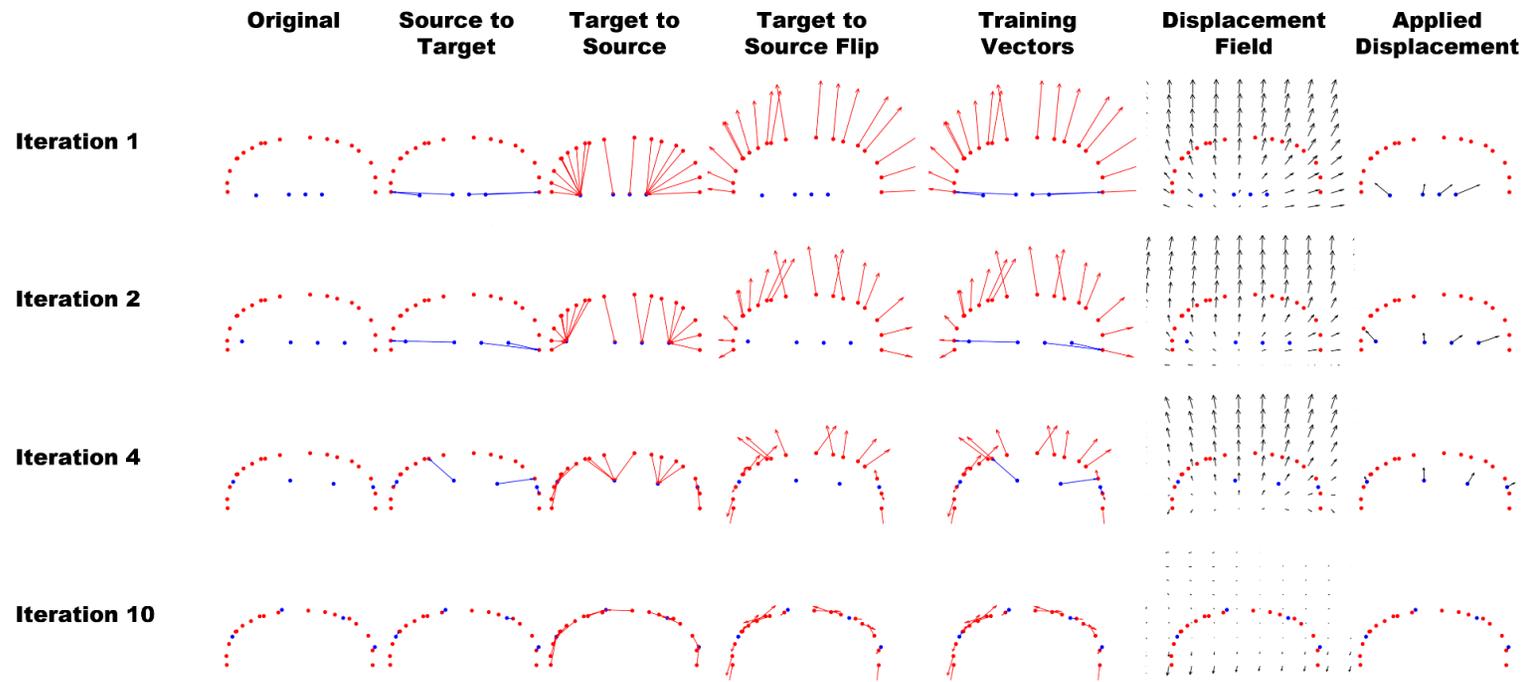

Figure 3: A 2D diagram showing the effects of smoothing on combining deformation between a source and target for morphing or between geometry 1 and geometry 2 in overclosure removal. Points are slowly morphed for each iteration to move the source nodes (blue) to the target nodes (red). The columns represent (Original) Target mesh (red nodes) and source mesh (blue nodes), (Source to Target) Nearest neighbor vectors from source mesh to target mesh, (Target to Source) Nearest neighbor vectors from target mesh to source mesh, (Target to Source Flip) Negated nearest neighbor vectors from target mesh to source mesh, (Training Vectors) Vectors used to train GRNN for desired displacement field ,(Displacement Field) Resulting displacement field from trained GRNN network for consistent smoothing level, (Applied Displacement) Resulting deformation vectors applied to source mesh from GRNN network with smoothing parameter. Black vectors represent direction and magnitude of nearest neighbors and resulting deformation vectors.

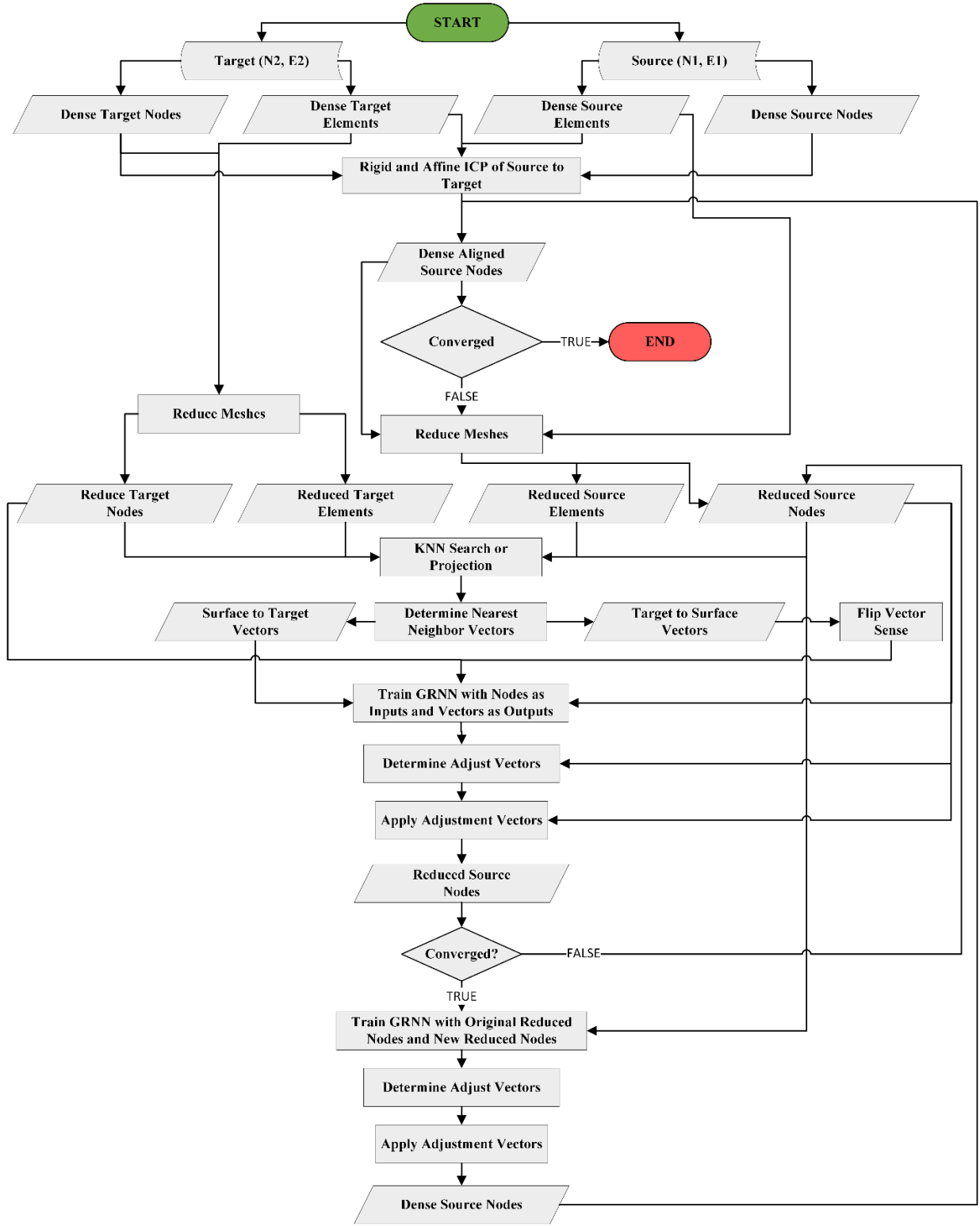

Figure 4: A flowchart showing the GRNN-morphing algorithm process for each pair of geometries.

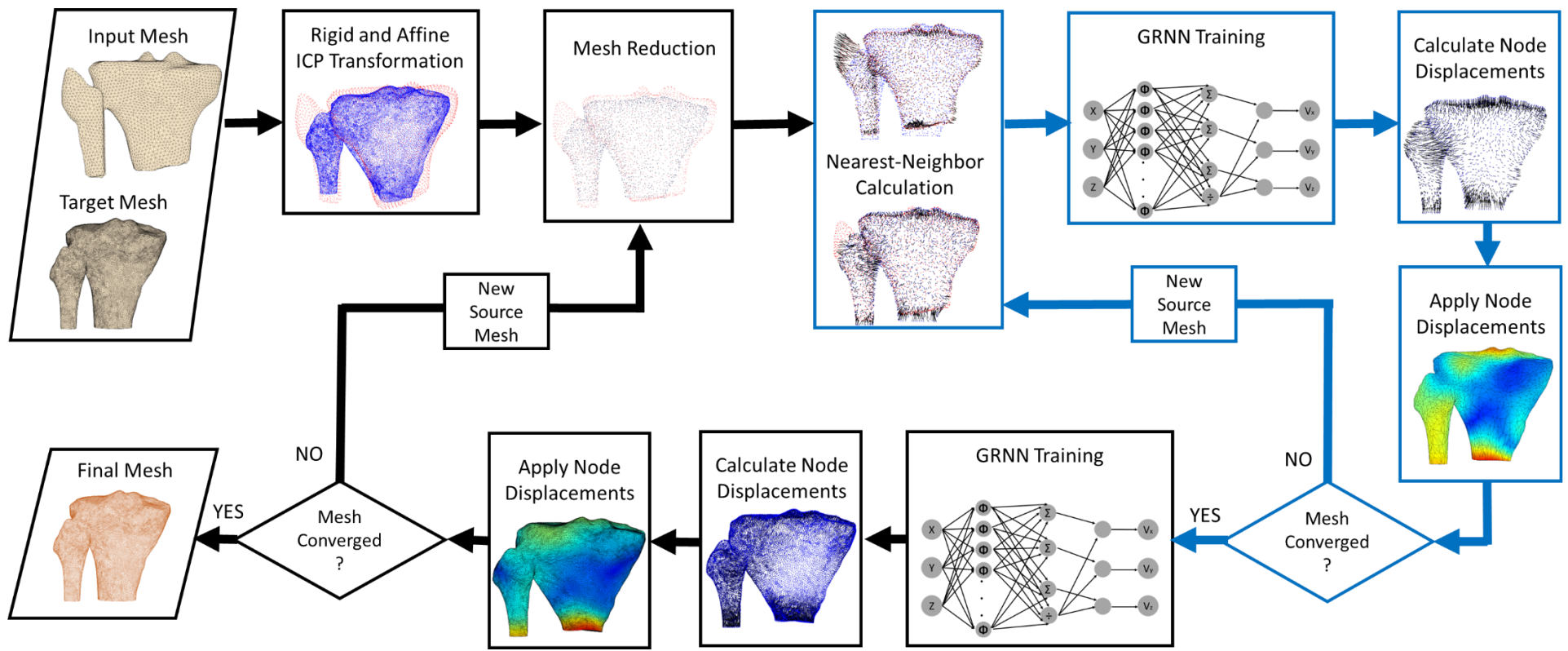

Figure 5: A diagram showing the steps applied to the original meshes of a combined tibia and fibula bone to morph the high density source to the low density target mesh. The inner loop represented in blue operates on the reduced meshes. When converged, adjustment vectors are trained and applied to the original mesh and checked for final convergence.

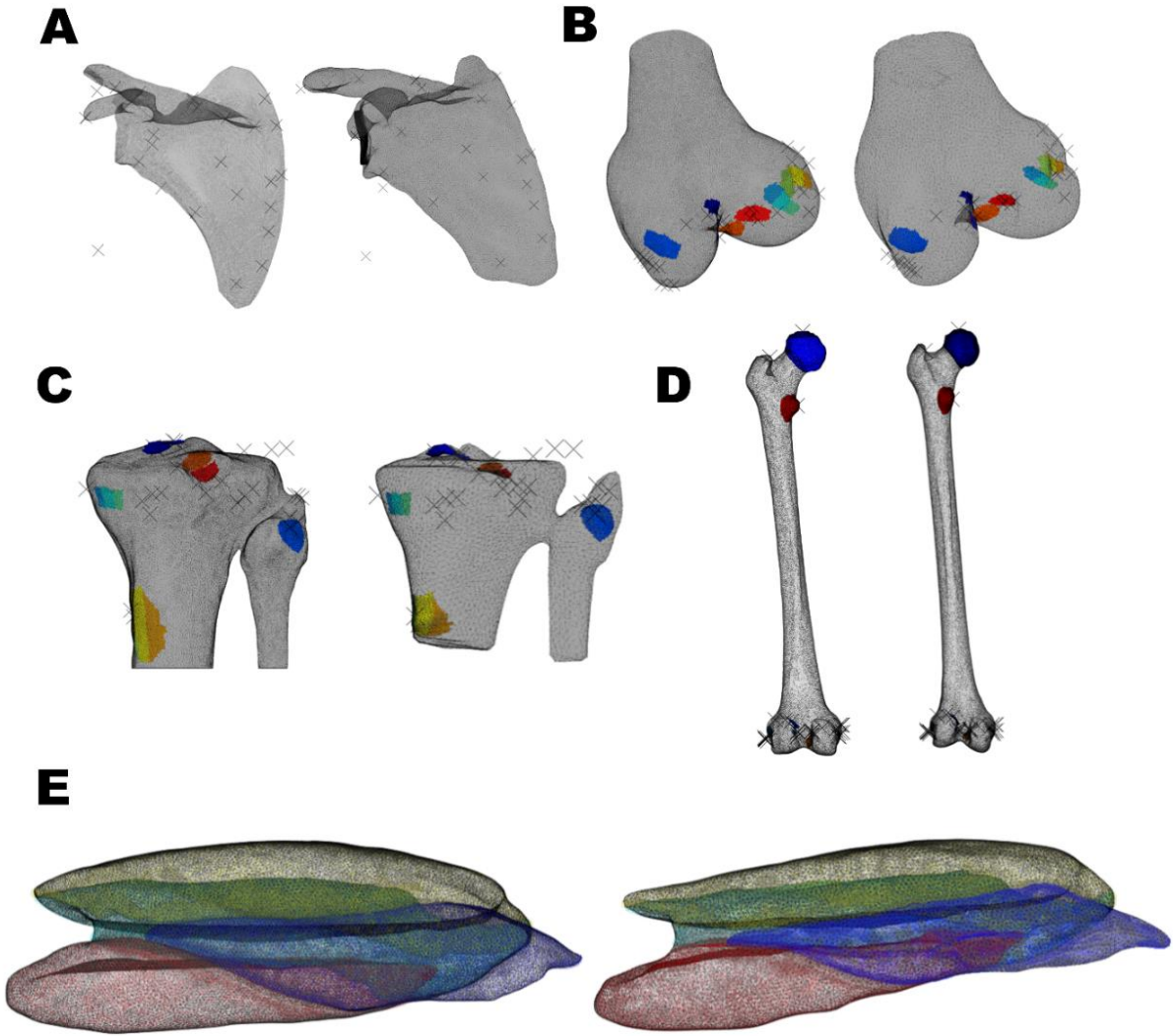

Figure 6: (A) Prediction of landmarks, attachment sites, and new geometry meshes from morphing geometries and predicted regions from morphing original locations to morphed geometries. (A) Scapula (B) Distal Femur (C) Proximal Tibia/Fibula geometry (D) Full Femur (E) Quadriceps Muscle Geometries. In all cases the original geometry with identified landmarks (Black X marks), attachments sites (color regions), and geometries (gray scale mesh) is shown on the left. The predicted geometries with predicted landmarks (Black X marks), attachments sites (color regions), and geometries (gray scale mesh) is shown on the right. For the predicted indvidual quadricep muscles in (E) the colors represent individual original and predicted muscle geometries for Vastus Lateralis, Vastus Medialis, Vastus Intermedius, and Rectus Femoris muscles.

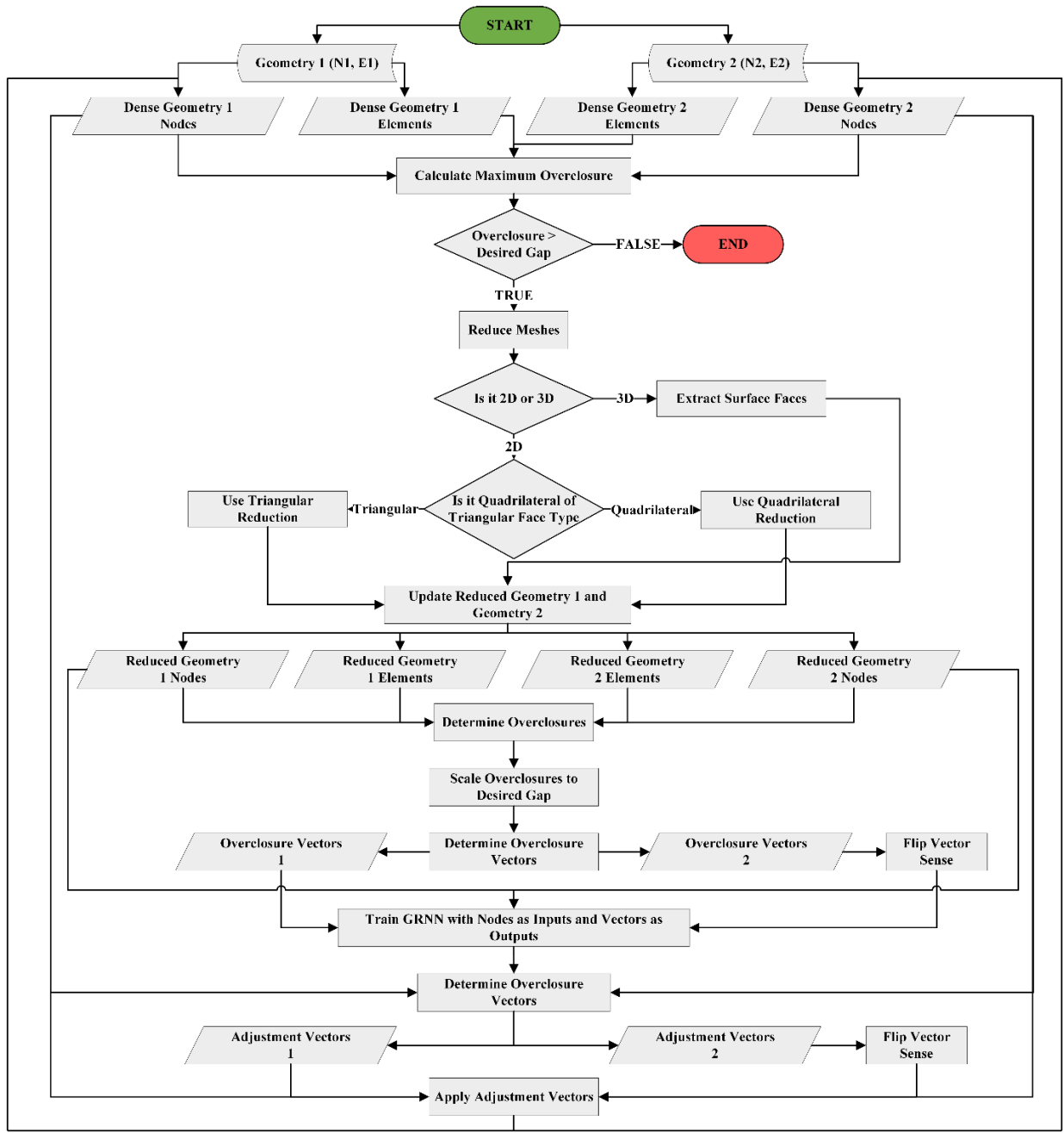

Figure 7: A flowchart showing the GRNN algorithm process for removing overclosures from a pair of 2D or 3D geometries.

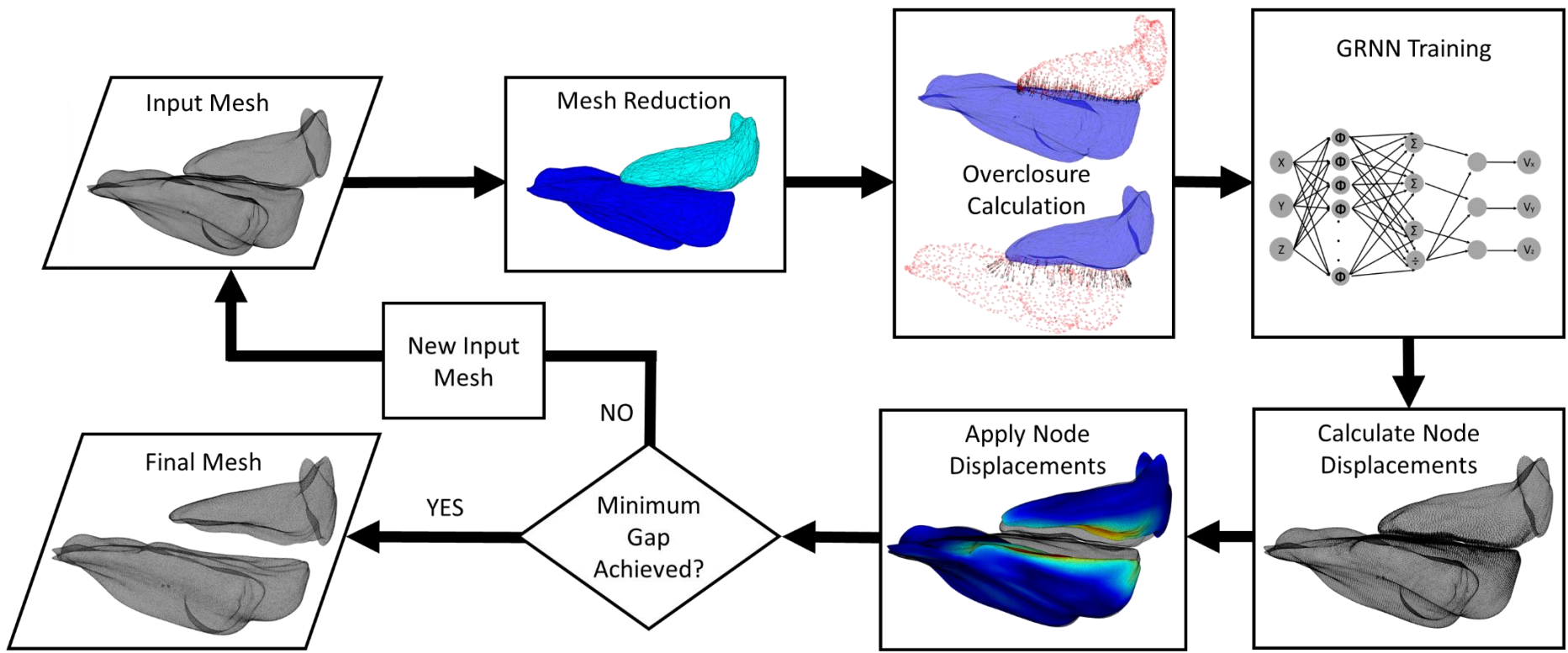

Figure 8: A diagram showing the steps applied to the original meshes of a gluteus maximus and gluteus medius muscle with an initial overclosure.

Andreassen et al. 38

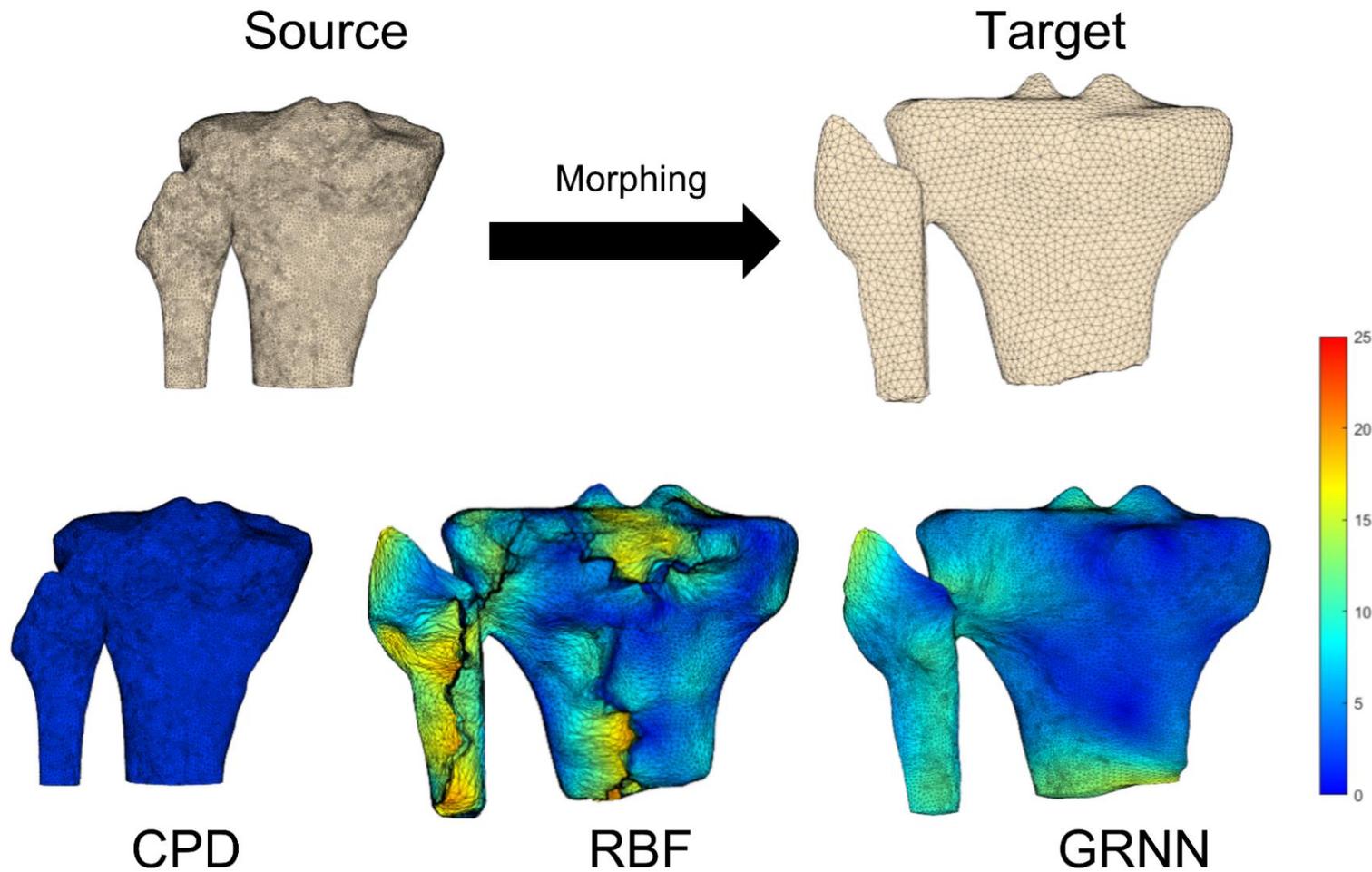

Figure 9: Source and target mesh geometries and resulting deformations for tibia fibula meshes for deformation of high density source to low density target mesh. Mesh morphed algorithms for CPD, RBF and GRNN algorithms show resulting meshes after the allocated time for morphing. Color gradients represent the distance traveled from the original node location.

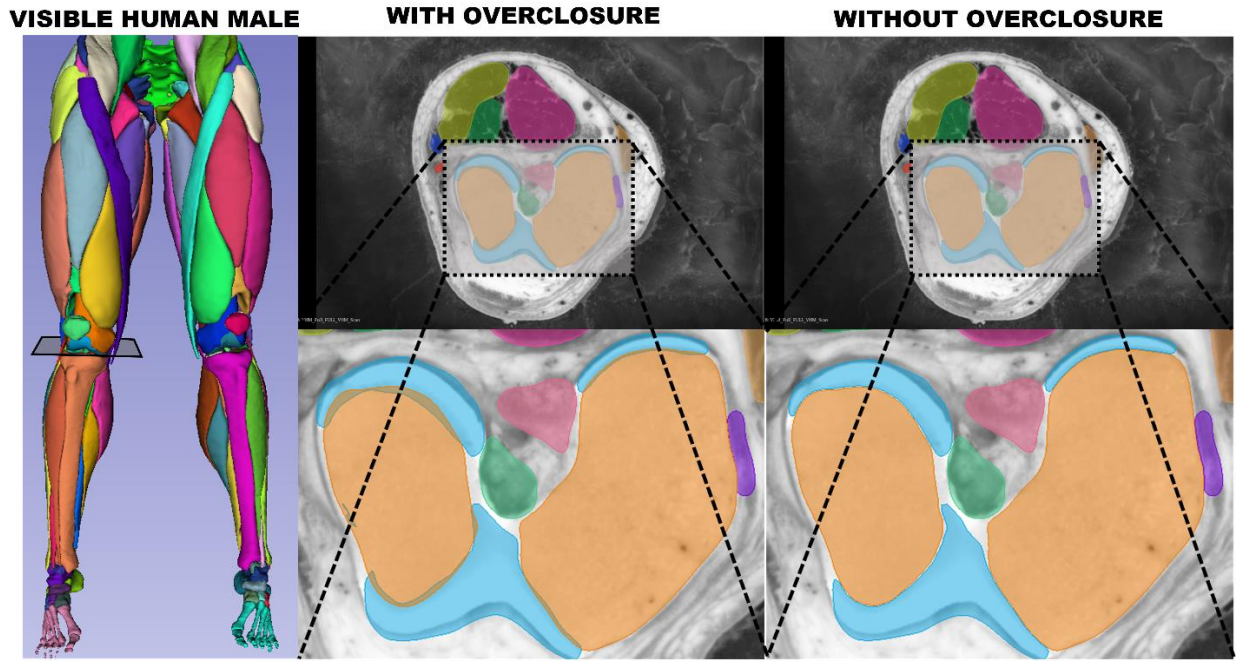

Figure 10: Visible human male geometry from Andreassen et al. The figure on the left shows the segmented geometries and indicates a slice of the segmented 3D geometries at the knee. The images in the middle show the geometries of the indicated slice after smoothing and remeshing of the original segmentation, highlighting the resulting overclosures between mating geometries of the knee. Images on the right show the final geometries of the indicated slice after the proposed RBF Network algorithm was used to remove all overclosures between mating geometries. In this case, the bones were left rigid, and all other geometries were deformed. Femur bone is shown in orange, and distal femoral cartilage is shown in blue.